\documentclass[prfluids,longbibliography,amsmath,amssymb,eprint]{revtex4-2} 
\usepackage{xcolor}
\usepackage{epsfig}

\usepackage[normalem]{ulem}

\def\l{\left}

\def\r{\right}

\def\Re{\mathrm{Re}}

\begin{document}

\title{Clogging a Porous Medium}
\author{H.~J. Seybold$^{1,2}$, Izael A. Lima$^{1,3}$ and Asc\^anio D. Ara\'ujo$^{1}$} 
\affiliation{$^{1}$Departamento de F\'isica,
  Universidade Federal do Cear\'a, Campus do Pici,
  60455-760 Fortaleza, Cear\'a, Brazil\\
  $^{2}$ETH Zurich, 8092 Zurich, Switzerland\\
  $^{3}$Centro de Forma\c c\~ao Antonino Freire - CFAF/UESPI, 64002-450 Teresina, Piau\'i, Brazil,}
  
\date{\today}

\begin{abstract}
  Flows through porous media can carry suspended and dissolved
  materials.  These sediments may deposit inside the pore-space and
  alter its geometry. In turn, the changing pore structure modifies
  the preferential flow paths, resulting in a strong coupling between
  structural modifications and transport characteristics.  Here, we
  compare two different models that lead to channel obstruction as a
  result of subsequent deposition. The first model randomly obstructs
  pore-throats across the porous medium, while in the second model the
  pore-throat with the highest flow rate is always obstructed first.
  By subsequently closing pores, we find that the breakdown of the
  permeability follows a power-law scaling, whose exponent depends on
  the obstruction model. The pressure jumps that occur during the
  obstruction process also follow a  power-law distribution with 
  the same universal scaling exponent as the avalanche size 
  distribution of invasion percolation, independent of the model. 
  This result suggests that the clogging processes and invasion 
  percolation may belong to the same universality class.
\end{abstract}
\pacs {}
\maketitle
\section{Introduction}
The clogging of porous media by deposition of dissolved or suspended
materials is a serious problem in many practically relevant
applications~\cite{baveye1998environmental,alem2015hydraulic,herzig1970flow}.
For instances, particle deposition considerably reduces the
permeability of porous filters and
catalysts~\cite{ochi1999two,reddi2000permeability} and has also a
major impact on the flow rate reduction in drainage layers
\cite{sansalone2012filtration}.  Deposition and calcification of
minerals in porous rocks is also important in other fields of
geoscience~\cite{mays2005hydrodynamic,noiriel2004investigation,mcdowell1986particle,shelton2016determining,upadhyay2015initial,osselin2016microfluidic}
and
carbon-sequestration~\cite{steefel2015reactive,cohen2015mechanisms,stack2015precipitation,soulaine2018pore}.
Due to its practical importance, the problem of clogging has
extensively been studied experimentally and theoretically since the
early
1980s~\cite{tien2013granular,lee2001network,vigneswaran1989experimental,salles1993deposition}.
A major challenge in understanding the clogging process is the dynamic
rearrangement of flow paths as a result of the changing pore geometry.
This interplay between geometry and flow controls macroscopic
properties such as the pressure drop required to maintain a certain
flow rate or the bulk permeability across the pore space.
\begin{figure}
       \centering  
       \includegraphics[width=0.85\columnwidth]{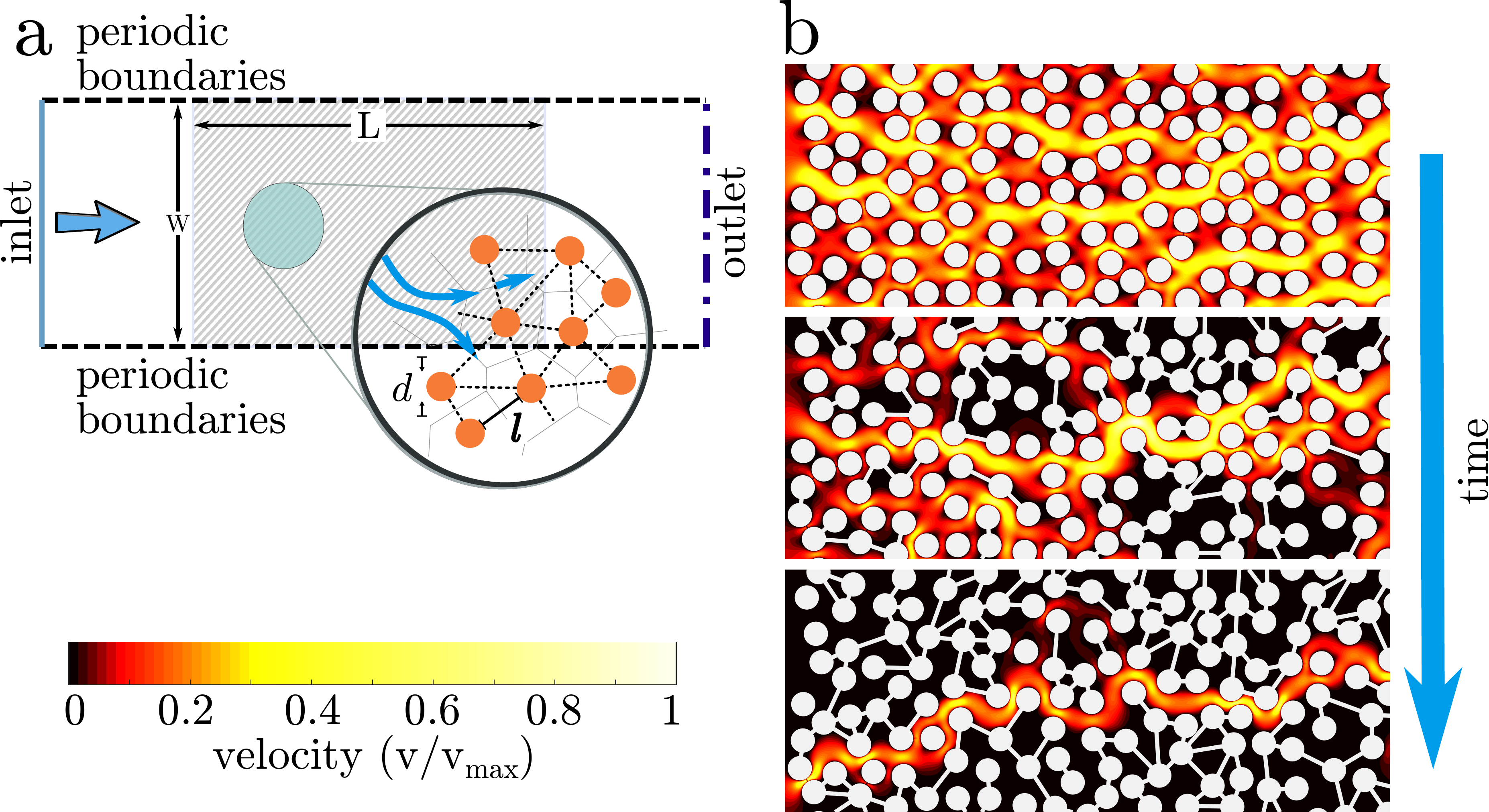}
       \caption{(a) Schematic view of our porous medium setup. The pore
         zone (hatched area of length $L$) consists of non-overlapping
         disks of a diameter $D$ (orange disks in the inset) placed at
         some distance $l$ between each other (inset). The connections
         between the different disks' centers form the edges of a
         Delaunay network (dashed) which can be successively
         obstructed.  Fixed inflow boundary conditions are applied at
         the inlet (left) and the outlet is defined by a vanishing
         pressure gradient. Periodic boundary conditions are applied
         at the top and bottom in order to reduce finite size effects.
         (b) Velocity magnitude for three snap shots taken at
         different times during the random clogging process. As more
         and more pores are closed the flow is getting more and more
         channelized.  The blue arrow indicate time evolution
         direction.  The porosity of the presented case was set to
         $\epsilon = 0.6$ and the inlet velocity was kept constant at
         $v_0=0.1$.
       } 
          \label{fig:fig_1}
\end{figure}
In most natural situations clogging occurs as a gradual process, where
pores subsequently narrow and finally completely close the flow path.
In order to simplify the modeling of this process, we make a quasi static
approximation, where the pores are obstructed instantaneously and the
fluid adjusts its flow path to the new conditions. Consequently, our
model consists of two alternating steps: First, an obstruction process
where a pore is closed based on an obstruction criterion, and second a
flow calculation every time a new obstruction is inserted.  

In analogy to filtration experiments small particles or solutes can be 
transported and trapped in any region of the porous medium,  while the 
transport of larger particles mainly occurs mainly in high flow regions, 
and thus pores with higher flow rates~\cite{mcdowell1986particle}.
Base on these observations, we analyze two different models to mimic
the obstruction process in a porous medium.  Our first model randomly
obstructs pore-throats across the porous medium, while in the second
model the pore-throat with the highest flow rate is always obstructed
first.

Although the two models show distinct differences in the initial clogging
stage, our results indicate that the final clogging transition just
before the porous medium is completely closed follows universal scaling
relations similar to those found for invasion percolation.

\section{Model and Simulations}

Natural porous media are typically disordered. Therefore, the porous
medium used for our analysis consists of circular obstacles with
non-dimensional diameter $D=1$. These obstacles have been randomly placed in a
rectangular domain of width $w=50$ and length $L=100$ measured in
units of obstacle diameters, Fig.\ref{fig:fig_1}.  To avoid overlap, 
the minimal distance between two obstacles' centers is set to $D+D/10$, 
corresponding to a minimal throat width of $D/10$.  
As the obstacles do not overlap, the porosity $\epsilon$ is controlled 
by the number of obstacles $n$ forming the pore zone:
$\epsilon=w\cdot L/ \left[n \cdot \pi (D/2)^2\right]$, Fig.~\ref{fig:fig_1}(a).

Fluid is injected in positive $x$-direction from the left to the right
and flow boundary conditions $\pmb{u}(0,y)=(v_0,0)$ are applied at the
inlet and zero pressure gradient $\nabla p=0$ at the outlet,
respectively (see Fig.~\ref{fig:fig_1}(a)).  Periodic boundary
conditions at the top and bottom of the square domain allow to reduce
finite size effects.  For simplicity we only consider an incompressible 
Newtonian fluid at low Reynolds numbers.  The Reynolds number
$\Re$ of our system is defined by $\Re=\varrho v_0 D/\mu$ where $\mu$
is the fluid's viscosity and $\varrho$ its density.

On the pore scale, the steady state velocity and pressure fields are
described by the incompressible Navier-Stokes equations
\begin{eqnarray} 
  \pmb{u}\cdot \nabla \pmb{u}
  &=&
      -\nabla p+\frac{1}{\mathrm{Re}} \nabla^ {2}\pmb{u}\\
  \nabla  \cdot  \pmb{u}&=&0,
\end{eqnarray} 
which can be solved numerically using a control volume finite
difference scheme~\cite{fluent2015ansys}. To avoid recreating the
numerical mesh every time a pore is obstructed, we included the
shortest connection between two neighboring obstacles as a ``internal
edge'', often called ``baffle'' into the meshing routine. These
internal edges are defined by the Delaunay triangulation network of
the obstacles' centers, Fig.~\ref{fig:fig_1}(a).  This trick highly
facilitates process of the opening and closing of throats by simply
changing the boundary conditions of the corresponding (internal) edge
from ``interior'' (no effect on the flow) to ``wall'' (no-slip solid
boundary on both sides of the edge).  After re-calculating the
pressure and velocity field, we determine the fluxes through the
different pore throats~\cite{araujo2006distribution}.  The width of a
pore throat $|\pmb{\ell}|$ is defined by the length of the Delaunay
edge between two neighboring obstacles minus two obstacle radii,
Fig.~\ref{fig:fig_1}(b) and the direction of $\pmb{\ell}$ is chosen
perpendicular to the Delaunay edge. Consequently the flux through each
throat can readily be calculated using
\begin{equation}
\phi=\int \pmb{v}\cdot \pmb{\ell},
\label{flux}
\end{equation}
were $\pmb{v}$ is the local velocity field calculated as the average
over the local velocity vector $\pmb{u}$ along the corresponding edge
$|\pmb{\ell}|$ of a pore throat.  Using the local fluxes, we then apply
our clogging rules by either randomly closing a Delaunay edge of the
obstacle triangulation or by selecting the edge with the highest flux.
The boundary condition of the corresponding edge in the numerical mesh
are changed from ``interior'' to ``wall'', which closes the throat
with an impermeable membrane.  After closing the throat we recalculate
the pressure and velocity field.  This process is then repeated
interactively until all flow paths connecting the inlet with the
outlet are closed. During this process we record local flow variables
like fluxes through the pore throats as well as the macroscopic quantities 
such as the pressure drop or permeability across the whole pore zone.

\section{Results and discussion}

Figure~\ref{fig:fig_1}(b) shows three different snapshot of a typical
simulation in which more and more pores are closed randomly (top to
bottom). The increased channelization caused by the successive closing
of pores is clearly visible, until only one single connected channel
is left, as we can see in Fig.~\ref{fig:fig_1}(b) (bottom).  

\begin{figure}
 \centering  
\includegraphics[width=0.75\columnwidth]{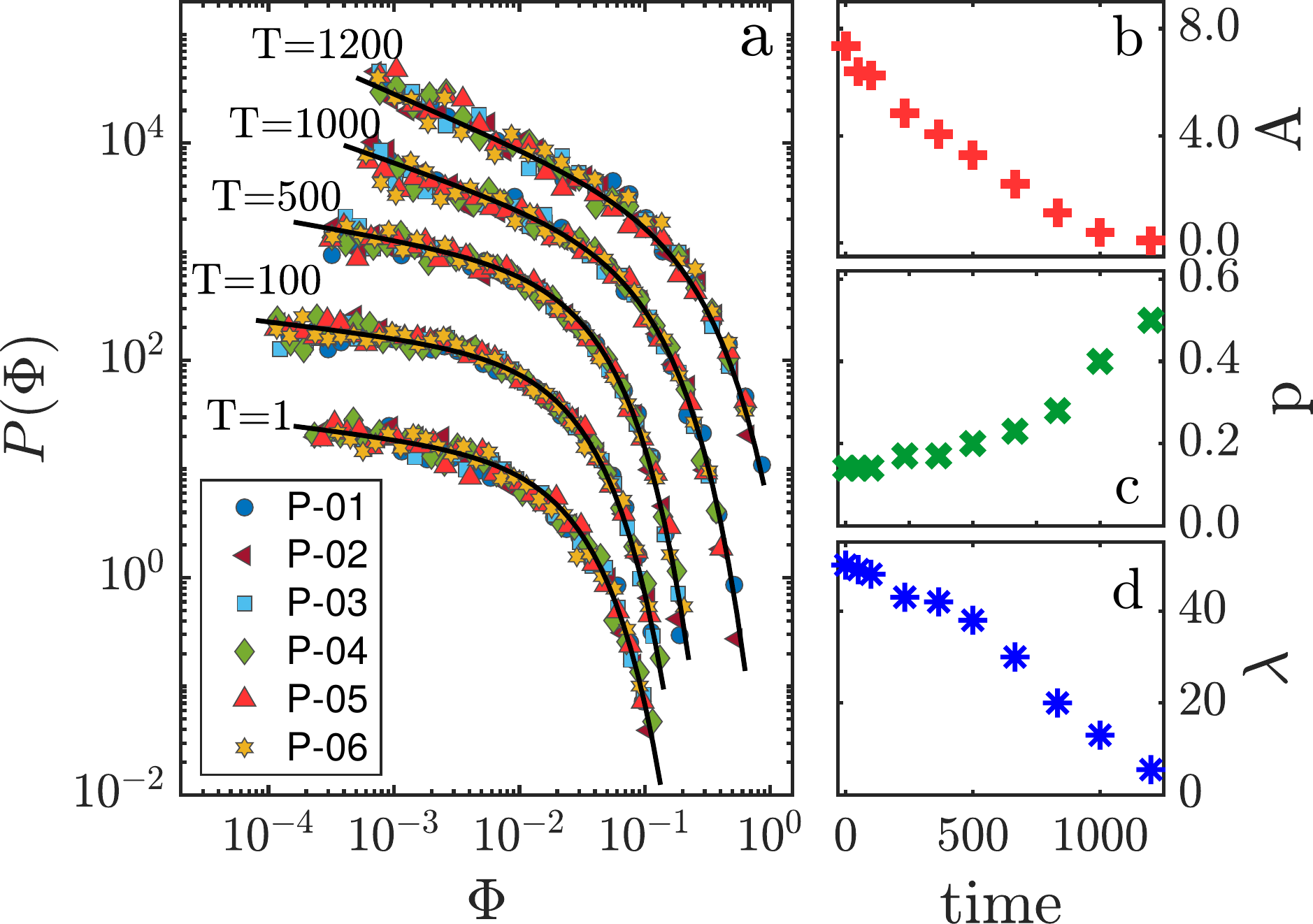}
\caption{(a) Logarithmic plot of the distributions of the normalized local
  fluxes $\Phi=\phi/\phi_{0}$ through the open pores at different time
  steps during the random obstruction. For better visibility each time
  plot is shifted vertically by one order of magnitude. The symbols
  correspond to different realizations of the porous medium with a
  constant porosity $\epsilon=0.6$. The solid
  black line is a non-linear fit using a scaling function
  Eq.(\ref{non_linear_fit}). The time evolution of the fit parameters 
   $A$, $p$ and $\lambda$ are shown as function of time $T$  in
  the left panels (b-d).}
  \label{fig:fig_2}
\end{figure}
In order to quantify the effect of preferencial channeling of the flow
during the clogging process, we calculate the flux $\phi$ through each
throat and determine their distribution $P(\Phi)$. Here
$\Phi=\phi/\phi_{0}$ are the normalized fluxes and
$\phi_{0}=v_{0}\cdot w$ is the total flux through the porous
medium. Fig.~\ref{fig:fig_2} shows $\Phi=\phi/\phi_{0}$ for the random
throat model at different times during the clogging process. For our
analysis we use six different realizations of the pore geometry with
the same porosity $\epsilon=0.6$ each represented by a different
marker in Fig.~\ref{fig:fig_2}(a).

During the clogging process, a systematic change of $P(\Phi)$ can be
observed.  Initially, the distribution $P(\Phi)$ has the shape of a
stretched exponential as reported by Ara\'ujo
{\it{et. al.}}~\cite{araujo2006distribution}.  However, with
progressive closing of pores a power-law component becomes more and
more apparent. Using a fit of the form
\begin{equation}
P(\Phi)=A~\Phi^{-p}~\exp({-\lambda \Phi})
\label{non_linear_fit}
\end{equation}
we can quantify this transition by looking at the evolution of 
the control parameters $A$, $p$ and $\lambda$ as shown in 
Fig.~\ref{fig:fig_2} (b-d).

For the first few hundred steps, the exponent $p$ that controls the
power-law regime, remains almost constant and then grows
monotonically. Conversely the parameter $\lambda$, approaches zero
leading to a more and more power-law shape distribution. Consequently,
close to the complete obstruction the flow is very heterogeneously
distributed in the porous medium with few active channels and
many stagnant zones of low or no flow, see Fig.~\ref{fig:fig_1}.
\begin{figure}[b]
  \includegraphics[width=0.5\columnwidth]{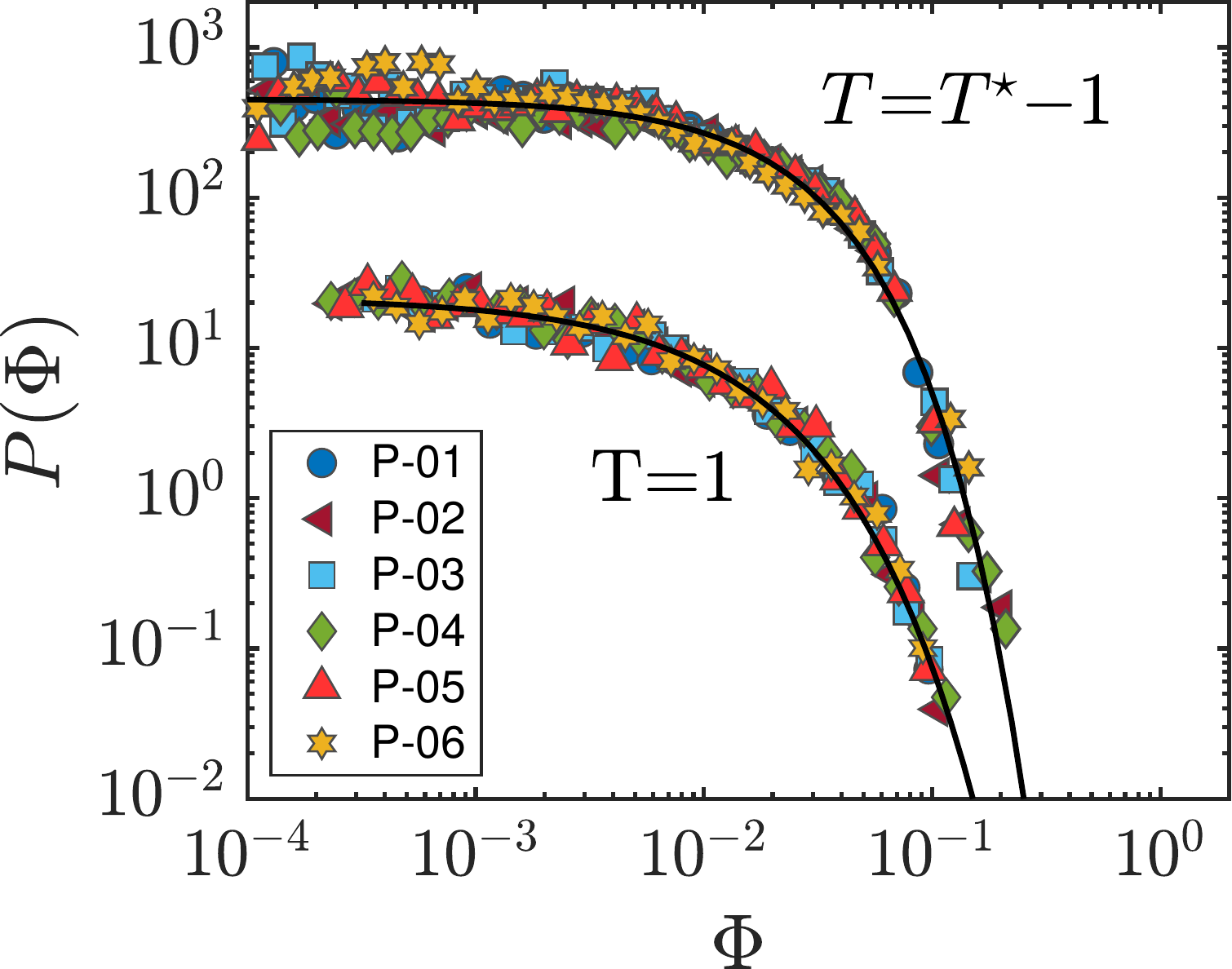}
  \caption{Plot of the normalized fluxes $\Phi=\phi/\phi_{0}$ through
    the open pores at the initial time $T=1$ and just before the
    porous medium is completely clogged $T=T^\star-1$ for the maximal
    flux obstruction model. Different from the random obstruction case
    the flux distribution seems to remain a stretched exponential
    (black solid line fit).}
  \label{fig:fig_3}
\end{figure}

Surprisingly, a similar behavior cannot be observed in the case, where
the pore-throats are obstructed according to the fluxes through the
throats. In this case, $P(\Phi)$ does not change over time and
preserves the original stretched exponential form during the whole
clogging process as can be seen in Fig.~\ref{fig:fig_3}.
%

We now draw our attention to the question how macroscopic quantities, 
such as the total permeability $K$ of the porous medium, are affected by the
successive closing of pores.

For low Reynolds numbers the flow rate $\pmb{q}$ through the pore zone
is a linear function of the applied pressure drop, 
\begin{equation}
\pmb{q}=\frac{K}{\mu}\nabla p
\end{equation}
where the proportionality constant $K$ is called
permeability~\cite{dullien2012porous,bear1972dynamics}.

\begin{figure}
  \centering  
  \includegraphics[width=0.98\columnwidth]{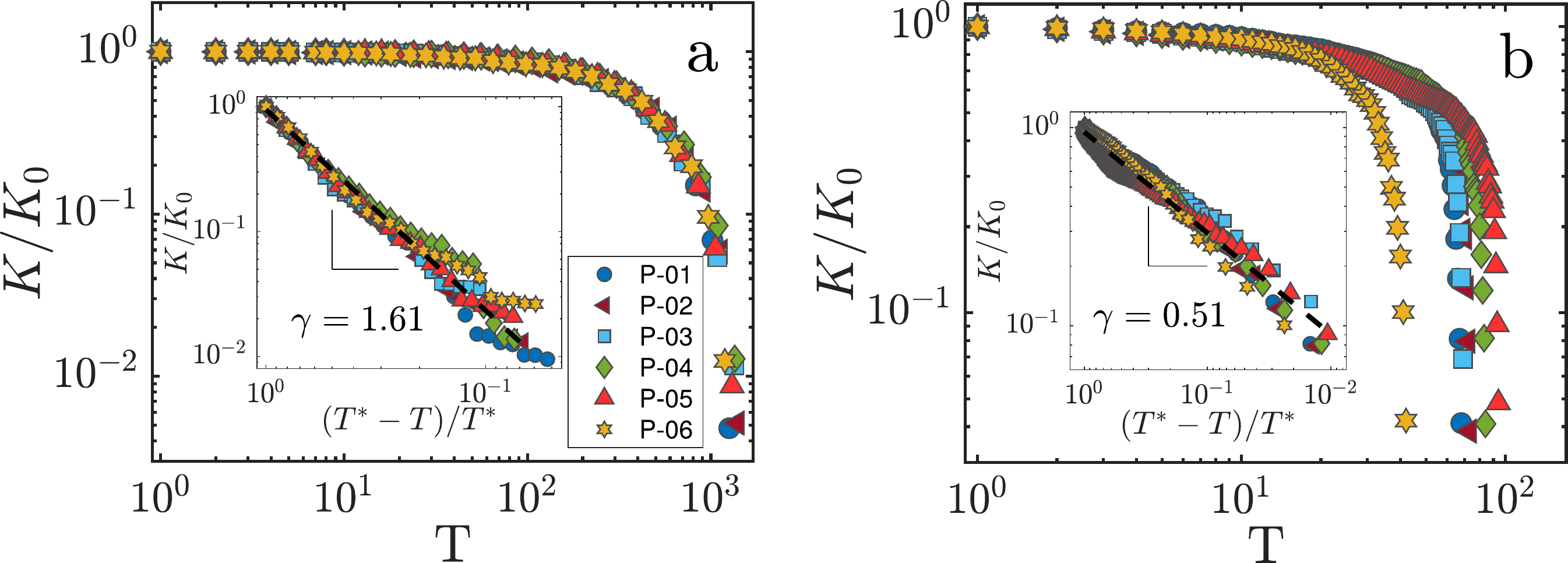}
  \caption{Log-log plot of the normalized permeability $K/K_{0}$ as a
    function of elapsed time. Symbols correspond to different
    realizations of a pore space with porosity $\epsilon = 0.6
    $. (a) Obstruction following the random model and (b) the
    maximal flux model.  The Reynolds number for all simulations was
    constant $\mathrm{Re}=0.001$. The insets in both figures show that
    the behavior of the permeability can be characterized by a power
    law
    ${K}/{K_{0}}\sim \l [ {(T^{\star}-T)}/{T^{\star}} \r]^{-\gamma}$,
    when centering and normalizing the timescale with the clogging
    time $T^\star$. }
  \label{fig:fig_4}
\end{figure}

Figure~\ref{fig:fig_4} shows the evolution of the normalized permeability
$K/K_{0}$ as a function of time $T$ for our two models, calculated for six
different realizations of the pore space (P-01 to P-06). The random
model is plotted in panel (a), while the max-flux model is shown in panel
(b); the parameter $K_{0}$, which is the initial permeability of the pore
zone before starting the obstruction process, was used to normalize
the different realizations. For both obstruction models two different
regimes can be observed. Initially, the normalized permeability
remains almost constant until reaching a point where it starts to
decay rapidly.

The exact time of the crossover depends on the details of the pore
geometry and the obstruction process and occurs roughly one order
of magnitude earlier in the max-flux case than in the random case.
Note, in the max-flux case the obstruction sequence is already
pre-determined by the pore geometry, while in the random case multiple
different sequences are possible starting from the same initial
configuration.
After the crossover, the normalized permeabilities decay fast until
the porous medium is completely blocked at $T=T^{\star}$(clogging
time).
While the main figure~\ref{fig:fig_4} shows the evolution of the
permeability from the beginning of the obstruction process, the insets
show the critical behavior when $K/K_{0}$ approaches the critical time
$T^{\star}$. By centering and normalizing time with $T^\star$, we can
compare the critical behavior of the two models around $T^\star$, even
though the maximum flow model clogs the porous medium much faster. The
critical breakdown of the permeability follows a power-law of the form
\begin{equation}
{K}/{K_{0}}\sim \l [ {(T^{\star}-T)}/{T^{\star}} \r]^{-\gamma},
\label{power_permea}
\end{equation} 
where the exponents which controls the power-law tail is
$\gamma=1.61\pm 0.02$ for the random model Fig.~\ref{fig:fig_4}(a) and
$\gamma=0.51\pm 0.03$ for the max-flux model
Fig.~\ref{fig:fig_4}(b). This attests that in the maximum flux model
the drop in permeability is more accentuated.

As we keep the flow rate constant during the obstruction process, the
pressure drop between the entrance and outlet of the pore zone is not
constant and increases over the course of the clogging process. The
increase of the pressure drop does not happen continuously, but in
bursts because some pores have a higher impact on the overall
permeability than others
\cite{jaeger2017mechanism,jager2017channelization} and thus lead to a
larger jump in the pressure drop across the pore zone.

\begin{figure}
       \centering  
	\includegraphics[width=0.8\columnwidth]{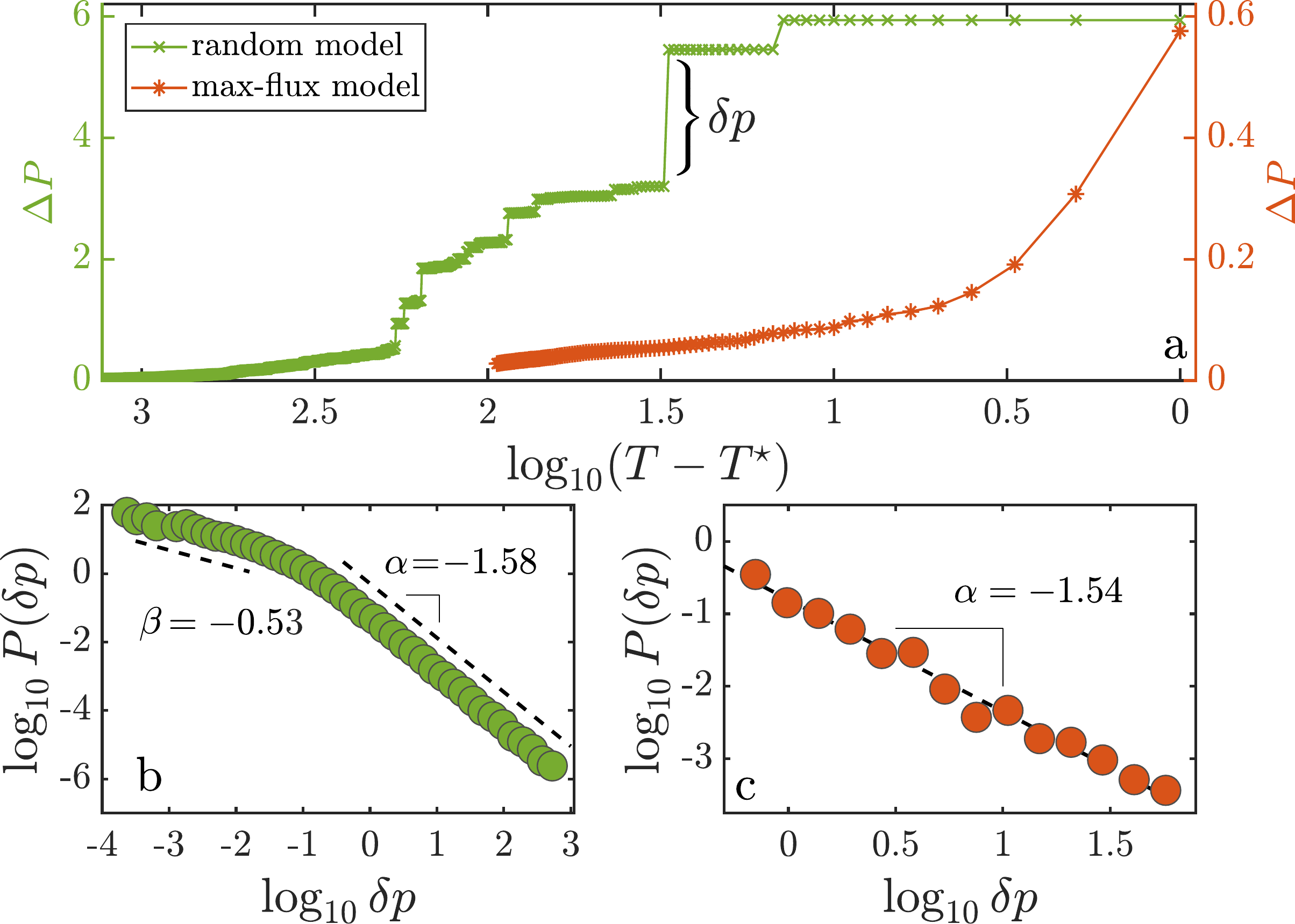}
        \caption{The pressure across the porous medium changes in
          discrete jumps $\delta p$ during the clogging process (inset
          panel a). The main panels show the distribution of pressure
          jumps for the random clogging model (a) and the maximal flux
          model (b). Note that both models show the same universal
          power law behavior close to the clogging transition, where
          the exponents are $\alpha=-1.58\pm 0.01$ and
          $\alpha=-1.54\pm 0.05$ respectively.}
     \label{fig:fig_5}
\end{figure}

Figure~\ref{fig:fig_5}(a) shows an example of how the pressure drop
across the pore zone changes over time for one sample of the random
obstruction process (green) and another sample of the max flux case
(orange). At the beginning of the clogging process the pressure bursts
tend to be smaller and increase in magnitude towards the clogging
transition.  While the evolution of the pressure drop is very
sensitive to the obstruction process and initial configuration, the
statistics of the bursts is independent of those details.

In order to quantify the pressure bursts during the clogging process,
we first calculate the size of the pressure jumps $\delta p$
(Fig.~\ref{fig:fig_5}(a)) for both obstruction model and then
determine their respective distributions.

Figure~\ref{fig:fig_5}(b) and (c) show the pressure jump
distributions for the random and the max flux case, respectively,
where the tails of both distributions follow a universal power-law
$P(\delta p)\sim( \delta p)^{\alpha}$ with similar exponents. More
precisely we find $\alpha=-1.54\pm0.05$ for the maximum flux model and
$\alpha=-1.58\pm0.01$ for random obstruction model. In the case of
random obstructions we also see a initial regime where the exponent
$\beta\approx -0.53\pm0.03$ is much smaller. This can be explained
as follows: During the initial stage of the obstruction the
possibility of closing an ``important'' channel is quite
low. However with successive closing of pores this probability
increases until at some point the random model behaves comparable to
the max flux model with a similar critical exponent.
It did not escape our attention, that the critical behavior of the
pressure jumps follows a scaling law which is very similar to the
avalanche size distribution of invasion percolation model (IP), namely
$\tau\approx -1.5$ \cite{roux1989temporal,araujo2004multiple}. This
behavior has also been confirmed
experimentally~\cite{crandall2009distribution}.
Similar to invasion percolation, the underlying dynamics of the
clogging is governed by extrema processes in a heterogeneous substrate
--- invading the capillary with the lowest pore pressure in IP or
closing the channel with the highest flux in our max-flux model. In
the case of random closing this process emerges only towards the end,
when most pores are closed and even random closing has a high chance
to select a pore with a extremal flux. This can be observed in
Fig.~\ref{fig:fig_5}(a) (green line), where towards the end of the
clogging process, the largest jumps are followed by periods of minimal
changes.
\section{Conclusion}
Our numerical study of the clogging process in a porous media has
revealed that the subsequent closing of pores displays many
characteristics of self organized criticality~\cite{bak2013nature},
including universal scaling laws. For the two different models of our
analysis, we found that the permeability breaks down following a power
law scaling. Specifically, we find scaling exponent of
$\gamma= 1.61\pm0.02$ for the random obstruction model and
$\gamma = 0.51\pm0.03$ when the pores with the highest flux is
obstructed first, respectively. This result may provide new insights
in the filtration as well in the drainage processes.
Although the scaling exponents for how the permeability approaches the
finite time singularity at $T=T^\star$ are different for the two
models, the distributions of the pressure jumps $\delta p$ which occurs
along with the successive closing of pores express a universal scaling
similar to the one observed in the avalanche distribution of invasion
percolation. The similarity in the exponents, together with
the analogy of extremal selection, suggests that invasion percolation
and the clogging of a porous medium may belong to the same
universality class which can help to improve the understanding of
the interplay between disordered pore substrate and the flowing fluid.

\begin{acknowledgments}
  The authors would like to thank the Brazilian agencies CNPq, CAPES
  and FUNCAP and the National Institute of Science and Technology for
  Complex Systems (INCT-SC) in Brazil and Petrobr{P\'a}s (``F\'isica do
  Petr\'oleo em Meios Porosos", Project Number: F0185) for financial
  support.
\end{acknowledgments}
\bibliographystyle{unsrtnat}
\bibliography{PhysRevFluids_v2.bib} 

\end{document}